\documentclass[twocolumn]{aastex63}  
\usepackage{graphicx,times,epsf}             %for PS/EPS graphics inclusion, new
\usepackage{longtable}
\usepackage{natbib}
\usepackage{amssymb}
\usepackage{amsmath}
\usepackage{psfig}
\usepackage{txfonts}
\usepackage[T1]{fontenc}
\shortauthors{Yang et al.}
\begin{document}

\title{An Empirical Bayesian Approach to Limb-darkening in Modeling WASP-121b Transit Light Curves}
\email{Fan Yang: sailoryf1222@gmail.com; sailoryf@nao.cas.cn}
\email{Ji-Feng Liu: jfliu@nao.cas.cn}
\author[0000-0002-6039-8212]{fan yang}
\affil{National Astronomical Observatories, Chinese Academy of Sciences, 20A
Datun Road, Chaoyang District, Beijing 100101, China\\}
\affil{IPAC, Caltech, KS 314-6, Pasadena, CA 91125, USA\\} 
\affil{Department of Astronomy, Beijing Normal University, Beijing 100875, People's Republic of China}
\affil{School of Astronomy and Space Science, University of Chinese Academy of Sciences,
Beijing 100049, China\\}

\author[0000-0002-8559-0067]{richard j. long}
\affil{Department of Astronomy, Tsinghua University, Beijing 100084, China\\}
\affil{Jodrell Bank Centre for Astrophysics, Department of Physics and Astronomy, The University of Manchester, Oxford Road, Manchester M13 9PL, UK\\}

\author {ji-feng liu}
\affil{National Astronomical Observatories, Chinese Academy of Sciences, 20A
     Datun Road, Chaoyang District, Beijing 100101, China\\}
\affil{School of Astronomy and Space Science, University of Chinese Academy of Sciences,
Beijing 100049, China\\}

\author[0000-0002-5744-2016]{su-su shan}
\affil{National Astronomical Observatories, Chinese Academy of Sciences, 20A
Datun Road, Chaoyang District, Beijing 100101, China\\}
\affil{School of Astronomy and Space Science, University of Chinese Academy of Sciences,
Beijing 100049, China\\}

\author {rui guo}
\affil{National Astronomical Observatories, Chinese Academy of Sciences, 20A
     Datun Road, Chaoyang District, Beijing 100101, China\\}
\affil{School of Astronomy and Space Science, University of Chinese Academy of Sciences,
  Beijing 100049, China\\}

\author[0000-0002-6434-7201] {bo zhang}
\affil{Department of Astronomy, Beijing Normal University, Beijing 100875,
People's Republic of China}

\author[0000-0002-5839-6744]{Tuan Yi}
\affil{Department of Astronomy, Xiamen University, Xiamen, Fujian 361005, P. R. China}

\author{Ling-Lin Zheng}
\affil{Department of Astronomy, Xiamen University, Xiamen, Fujian 361005, P. R. China}

\author{Zhi-Chao Zhao}
\affil{Department of Astronomy, Beijing Normal University, Beijing 100875,
People's Republic of China}

\begin{abstract}
We present a novel, iterative method using an empirical Bayesian approach for modeling the limb darkened WASP-121b transit from the TESS light curve. Our method is motivated by the need to improve $R_{p}/R_{\ast}$ estimates for exoplanet atmosphere modeling, and is particularly effective with the limb darkening (LD) quadratic law requiring no prior central value from stellar atmospheric models. With the non-linear LD law, the method has all the advantages of not needing atmospheric models but does not converge. The iterative method gives a different $R_{p}/R_{\ast}$ for WASP-121b at a significance level of 1$\sigma$ when compared with existing non-iterative methods. To assess the origins and implications of this difference, we generate and analyze light curves with known values of the limb darkening coefficients (LDCs). We find that non-iterative modeling with LDC priors from stellar atmospheric models results in an inconsistent $R_{p}/R_{\ast}$ at 1.5$\sigma$ level when the known LDC values are as those previously found when modeling real data by the iterative method. In contrast, the LDC values from the iterative modeling yields the correct value of $R_{p}/R_{\ast}$ to within 0.25$\sigma$. For more general cases with different known inputs, Monte Carlo simulations show that the iterative method obtains unbiased LDCs and correct $R_{p}/R_{\ast}$ to within a significance level of 0.3$\sigma$. Biased LDC priors can cause biased LDC posteriors and lead to bias in the $R_{p}/R_{\ast}$ of up to 0.82$\%$, 2.5$\sigma$ for the quadratic law and 0.32$\%$, 1.0$\sigma$ for the non-linear law. Our improvement in $R_{p}/R_{\ast}$ estimation is important when analyzing exoplanet atmospheres.
\end{abstract}
\keywords{Exoplanet atmospheres(487)---Exoplanet detection methods(489)---Transit photometry(1709)---Exoplanet systems(484)}

\section{introduction}

Stellar limb darkening is an optical effect caused by photons originating shallower inside the star as distances from the center of a star increase. As a consequence, stellar surface brightness decreases with increasing radius. Being able to represent surface brightness and limb darkening accurately is fundamental to exoplanet transit modeling. Typically, limb darkening is characterized by what is referred to as "limb darkening laws". The most popular laws are the linear, quadratic, and non-linear limb darkening laws \citep{claret2000,claret2011}. Deriving limb darkening coefficients (LDCs) by finding adjustments from stellar atmosphere spherical modeling has been proved effective even though it sometimes induces biases for exoplanet parameters, e.g. to the radius ratio of exoplanet to star ($R_{p}/R_{\ast}$) 
\citep{Espinoza2015}. Removing or at least reducing this potential uncertainty in
$R_{p}/R_{\ast}$ caused by limb darkening modeling is important when constraining the exoplanet atmosphere through precise measurements of bandpass dependent $R_{p}/R_{\ast}$ \citep{Deming2005, Karkoschka2011, Madhusudhan2019}.

In recent applications of exoplanet transit fitting, LD has been implemented with floating coefficients (LDCs) with a Gaussian prior based on theoretical stellar atmosphere models \citep{Nikolov2018, Evans2018, Demangeon}. This approach is a compromise as a result of quality limited data and stellar atmospheric model approximations. Fully free LDCs in fitting cause a large uncertainty in the estimation of LDCs and this uncertainty propagates to the other transit parameters \citep{Cabrera2010,borde2010}. 
Applying LDCs from an independent method can significantly reduce the uncertainty. Unfortunately, LDCs are difficult to obtain accurately. Direct observation is only available when host stars are close enough to be spatially resolved \citep{Haubois2009}.
Microlensing is another independent method to measure the stellar limb-darkening \citep{Witt1995, Dominik2004, Zub2011}. Theoretical predictions from stellar atmospheric models sometimes show inconsistencies in their inputs \citep[e.g., PHOENIX and ATLAS;][]{PHOENIX, ATLAS}. These inconsistencies are reported to cause significant biases to $R_{p}/R_{\ast}$ at a level of 1-10 percent when fixing LDCs during transit fitting\citep{Csizmadia2013, Espinoza2015}.

\citet{Howarth2011} shows that the LDCs obtained from stellar atmosphere models and transit models are sometimes different due to the different fitting processes (synthetic-photometry/atmosphere-model effects). More interestingly, the difference yields approximately constant values for the same planet (within reasonable stellar parameter ranges). The discussion about geometry is described in, for example, \citet{Espinoza2015}. The offset due to geometry is also nearly constant but only for 3500 K $\le$ $T_{eff}$ $\le$ 7500 K
\citep[see Figure 6 of][]{Espinoza2015}.
%The discrepancy due to input physics was previously reported \citep{Kurucz1979, Husser2013}.

Another unsolved issue is the standard deviation ($\sigma$) of the LDC prior which is usually set as a Gaussian distribution \citep[e.g.][]{silvotti2014, Chen2018, Luque2019}. The $\sigma$ should take into account knowledge obtained from stellar modeling predicted LDCs, the transit fitting experience, and observational data. The selected value of $\sigma$ is quite different among different works \citep{Wang2013, Siverd2018, Wang2019, Barkaoui, Shporer2020}. The LDC prior significantly affects the final LDC value for transit fitting.

The LD error propagation to $R_{p}/R_{\ast}$ is reported to be sometimes up to 1$\%$ when applying theoretical LDCs as a prior \citep{muller2013, Espinoza2015}. This $R_{p}/R_{\ast}$ uncertainty is not negligible when analyzing exoplanet atmospheres through wavelength dependent $R_{p}/R_{\ast}$ \citep{seager2010,Madhusudhan2019}. For atmosphere detected sources, $R_{p}/R_{\ast}$ varies in different observational bands from a few tenths of a percent to a few percent, correlated with the width of the bands. The observational significance level ranges from less than one to a few $\sigma$ \citep{Charbonneau2002, sing2016, Evans2017}. It has been reported that the bias can be reduced but not fully prevented when applying higher-order LD laws, or using higher precision photometry, e.g., Kepler, Transiting Exoplanet Survey Satellite (TESS) and, once launched, the James Web Space Telescope (JWST) \citep{muller2013, Csizmadia2013, Espinoza2015}. The creation of unbiased $R_{p}/R_{\ast}$ modeling is still in the explorative stage.

Some pioneering works have been performed applying stellar atmosphere model independent LDC priors, using Kepler light curves and simulation light curves \citep{muller2013, Csizmadia2013}. They obtain the $R_{p}/R_{\ast}$ bias caused by different synthetic stellar atmosphere models and compare the result with parameters derived with free LDCs. A transit light curve with the signal-to-noise ratio (SNR) $\sim$10 is reported to cause a 2$\%$ uncertainty to $R_{p}/R_{\ast}$ when applying a fully free LDC prior \citep{Csizmadia2013}.

Facing the same challenge in the TESS era, we present an iterative method in framework of empirical Bayes \citep{EmpiricalBayes} to obtain LDCs without resorting to a synthetic stellar model and perform a proof of concept assessment. The paper is organized as follows. Section 2 describes the iterative method and its application to the light curves of WASP-121b in the TESS survey. In section 3, we perform Monte Carlo simulations to assess the transit parameters derived. We examine the biases of transit parameters caused by biased LDC priors when applying classic (hereafter termed non-iterative) methods. Any multimodality of LDCs is addressed in this section. We also examine whether there is any evidence of overfitting for both the quadratic and non-linear law. Section 4 summarizes our findings, particularly for exoplanet atmosphere modeling.

\section{Iterative Method for Limb Darkening With TESS Light Curves}

Limb darkening interacts with other transit parameters in transit fitting \citep{Mandel_Agol2002}. In this section, we describe an iterative method for implementing LD and apply it to WASP-121b. Also, we present non-iterative fittings applying float LDCs with Gaussian priors to WASP-121b and a TESS identified source HD-219666b for comparison.

The discovery of exoplanet WASP-121b was reported by \citet{Delrez}. 
The host star has a mass of $1.353^{+0.080}_{-0.079}M_{\odot}$, a radius of $1.458\pm0.030R_{\odot}$ and luminosity of 10.4 mag in V band. The planet is a hot Jupiter with a period of 1.28 days, a mass
of $1.183^{+0.064}_{-0.062}$ Jupiter mass ($M_{J}$) and a radius of $1.865\pm0.044$ Jupiter radius ($R_{J}$).

Exoplanet HD-219666b was discovered by \citet{Esposito2019} using TESS data. It is a hot Neptune around a G7 star (mass $0.92\pm0.03M_{\odot}$, radius $1.03\pm0.03R_{\odot}$) with a period of 6.04 days. The planet has a mass of $16.6\pm1.3$ Earch mass ($M_{\oplus}$) and a radius of $4.71\pm0.17$ Earth radius ($R_{\oplus}$).

\subsection{Mathematical justification of the iterative method: empirical Bayes}

Bayes theorem is used to infer parameters from our preknowledge and the observations. Bayesian inference \citep{Feroz2009} can be expressed as 
\begin{equation} P(\mathbf{\Theta}|\mathbf{D}, M) =
\frac{P(\mathbf{D}|\,\mathbf{\Theta},M)P(\mathbf{\Theta}|M)}
{P(\mathbf{D}|M)},
\label{BI}
\end{equation}
where $\Theta$ stands for parameters, $\mathbf{D}$ for the data, and $\mathbf{M}$ for the model. $P(\mathbf{\Theta}|\mathbf{D}, M)$ is the posterior distribution, $P(\mathbf{D}|\,\mathbf{\Theta},M)$ the likelihood, $P(\mathbf{\Theta}|M)$ the prior, $P(\mathbf{D}|M)$ the Bayesian evidence.

%For instance, the $\sigma$ of the Gaussian prior if applied, is not always taken as the observational uncertainty in former observation. It should include the weight of former observation which is an important and difficult issue in Bayes theory\citep{efron_2010}. 
%This causes the fact that in planet transit fitting, the $\sigma$ of LDC prior is somewhat arbitrary \citep{Shporer2020}. } 

The model ($M$) contains the Bayesian idea of theories, experience, the weight towards former and new observations. The prior is a model prediction that takes the above into account \citep{efron_2010}. The prior shapes the form of the posterior distribution. However, a good model-based prior is sometimes hard to obtain, e.g., the arbitrary choice of $\sigma$ in the LDC prior in planet fitting \citep{Shporer2020}. The empirical Bayes method \citep{MorrisEmBa, EmpiricalBayes} solves this issue, in a general way, obtaining the prior from both the new data and the former knowledge.

The external knowledge is treated as former observations $\mathbf{D}_{1}$, $\mathbf{D}_{2}$, ..., $\mathbf{D}_{m}$, with the new observations as $\mathbf{D}_{m+1}$, $\mathbf{D}_{m+2}$, ..., $\mathbf{D}_{n}$. The new observations here refer to different datasets from either different observing runs or just resampling of one sample, e.g., jackknife or bootstrap sampling. A basic assumption is that the relation between the observations and the parameters $\mathbf{\Theta}$ is the same among $\mathbf{D}_{1}$, $\mathbf{D}_{2}$,..., $\mathbf{D}_{n}$. If we use the prior obtained purely from $\mathbf{D}_{1}$, $\mathbf{D}_{2}$, ..., $\mathbf{D}_{m}$, empirical Bayes is the same as non-empirical Bayes.

Empirical Bayes allows a prior distribution form to be assumed from former observations and obtains the parameters of the distribution form through both the former observations and the new observations (data). The prior obtained is finally applied in Bayes inference (Equation \ref{BI}). Empirical Bayes has recently become a widely used method for Bayes inference \citep{EP2001,EP2004,EP2007}.
The conditions under which we use the method are similar to those in other astronomical work, for example \citet{EBA2018} and \citet{EBA2020}.

In our approach, we apply an empirical Gaussian prior with the $\sigma$ value from foreknowledge, and iteratively fit the data to find the proper prior center for limb darkening coefficients.
Iteration is commonly used with empirical Bayes \citep{efron_2010}.
%, and has become one of the basic and most important methods in machine-learning.
It is more mathematically sound to use subsamples generated from the observational data in the iterative process, in considering if the datasets were not identically distributed and if there were any abnormal data points. The subsample here can be obtained by a resampling method, e.g., jackknife or bootstrap. In each iteration cycle, the subsample is updated and the model parameters inherit from the previous iteration. The subsample can be replaced by the whole sample if the difference between using a subsample and the whole sample not evident.
The rationale behind this is that reuse of the whole dataset is being treated as the use of an identical distribution. In this work, the subsample is replaced by the whole dataset. We have tested the subsample approach by randomly omitting 10$\%$ of the whole data set in each iteration, and found only negligible differences from just using the whole data set for all iterations. In addition, the resampling is not with much practical significance in our empirical Bayes approach. Because it would be impossible for us to improve the precision of transit parameters at the level of 1 $\sim$ 2 $\sigma$ if the datasets were not identically distributed.
%The meaning behind this is the whole dataset can be treated as an identical distribution.
%We test to generate the subsample by omitting 10$\%$ of the whole data during the iteration and find a negligible difference if using the whole data during the iteration.
\subsection{Iterative method of limb darkening determination} 
%In subsection headings, only the first word, in MNRAS articles, should have an upper case first letter. All the other words should be in lower case.

A parameterized transit model is typically used \citep{Mandel_Agol2002} to determine LD by fitting transit light curves. An alternative choice proposed recently is through tabulated stellar intensities \citep{Short2020}. Monte-Carlo Markov Chain methods (MCMC) and multimodal nested sampling algorithms (MULTINEST) have proved to be effective in determining multi-dimensional distribution of parameters \citep{pymc, EMCEE, Feroz2008, Feroz2009}. We obtain our fitting results mainly from an MCMC routine \citep[PYMC;][]{pymc} and find the results are similar when we replace PYMC with a MULTINEST routine \citep[PyMultiNest;][]{pymultinest}. 

Our method is particularly effective for the case when the stellar atmosphere model predicted LD is significantly different from the LD obtained from transit fitting. The difference could come from the systematic errors in the stellar models or the different fitting processes \citep{Howarth2011}. It is clear from \citet{Howarth2011} that offsets between the LDC values from different modeling schemes are approximately constant for the same planet in different wavelengths (see their Figures 6 and 7). Also, the LDC offsets from different models are almost constant or weakly dependent on the stellar effective temperature. The offset values for different planets are related to, e.g., the impact factor of the planet system, the temperature, and the gravity of the star. This nearly constant offset acts to support an approach that knowledge of LD uncertainty from stellar atmospheric models is also helpful for setting the LD prior $\sigma$ during transit fitting. The widths of the distributions inferred from the model grids should however not be severely affected.

Our method iterates the MCMC fitting process to the light curves, rather than just taking the result of a single MCMC fitting process. In every iteration loop, the LDC priors updates with the fit result from the previous loop. Every loop is an independent MCMC transit fitting process and does not differ from the classic transit fitting as described in Sec. 2.3. For the avoidance of doubt, no MCMC chain-breaking takes place.

LDCs start with arbitrary initial priors. We determine LDCs mean values from the morphology of the transit light curves. The limb darkening output from light curve fitting is set as the input prior to the next iteration. For the most general case, it would be possible to define the iteration step as some fraction of the difference between two iterations. We currently do not use that level of generality and our fraction is 100 percent. A theoretical discussion on the choice of the iteration step is beyond the scope of this work.

%Our LDC prior for each parameter is a separate Gaussian with a $\sigma$ of 0.05. In non-iterative fitting, any value of $\sigma$ is a compromise between the fixed LDCs from the stellar atmosphere and a fully free LDCs \citep{Shporer2020}. Too narrow a prior over-emphasizes the importance of a stellar atmosphere model and could yield biased fitting results. Too broad a prior causes a large uncertainty in the fitting parameters. The choice of $\sigma$ is empirical and somewhat arbitrary \citep{Shporer2020}. For iterative fitting, $\sigma$ is not as important as for the non-iterative method. We adopt a popular value of 0.05 as in the work TESS exoplanet discovery of \citet{Wang2019}. The appropriateness of $\sigma$ is tested by simulation that the posterior of LDCs in a single MCMC fitting represents the distribution of LDC estimations from 1000 simulated light curves (in Sec. 3.3). 

Our LDC prior for each parameter is a separate Gaussian with a $\sigma$ of 0.05. A Bayesian interpretation of the prior is a mean value with a $\sigma$ of 0.05, predicted by the fitting model which includes, e.g., stellar atmospheric knowledge, the transit fitting experience, and the TESS data knowledge. The resultant effect is very hard to quantify and causes difficulties in LDC prior $\sigma$ applications \citep{Shporer2020}. Utilizing the "control variate method" to examine the relative influences of stellar atmospheric knowledge, the transit fitting experience, and the TESS data knowledge would shed light on the quantitative prior $\sigma$ choice. However, it is very hard to achieve and beyond the scope of this work. We apply an empirical LD prior $\sigma$ taken from similar works. The $\sigma$ value of 0.05 is empirically used in TESS exoplanet fitting \citep{Wang2019}. For the non-iterative method, the prior central value is mainly derived from stellar atmospheric models. For the iterative method, the prior center is obtained from the new observations.  
%The appropriateness of $\sigma$ is partially tested by simulation that the posterior of LDCs in a single MCMC fitting represents the distribution of LDC estimations from 1000 simulated light curves (in Sec. 3.3).
%This idea is also consistent with the ideas in \citet{Howarth2011}. The LDCs of the same planet may introduce a simple offset to the whole grid of limb darkening coefficients. The widths of the distributions inferred from the model grids should however not be severely affected. 

The fitting is repeated ten times with the same prior and we take the median value as the LDC output, to reduce fluctuations. We repeat the loop if the standard deviation of LDCs among duplicated fittings is larger than 0.05. We also repeat the iterative loop when the difference in the LDCs compared to the previous loop is larger than 0.05. These actions improve the convergence speed.

\subsection{Light curves derived from TESS photometry}

The Transiting Exoplanet Survey Satellite \citep[TESS;][]{Ricker2015} is launched in 2018, aiming to discover transiting exoplanets in the solar neighborhood. TESS employs four cameras with a total field of view (FOV) 24$\times$96 square degrees. TESS captures 30 minutes cadence images, named Full Frame Image, for all the sources in the FOV, and 2 minutes images as well as photometry products for certain sources. Photometry precision reaches $\sim$1$\%$ at 16 magnitudes in a broad optical band ($0.6-1$\,$\mu$m).

We use 2 minutes cadence image frames, named Target Pixel File (TPF), to generate light curves. The pipeline versions are spoc-3.3.57-20190215 for WASP-121b, and spoc-3.3.36-20180925 for HD-219666b. A comprehensive description of the data reduction is presented in earlier work (Yang et al. 2019, submitted). We briefly overview the reduction here. Also, we have compared our result with the light curve generated by TESS Science Processing Operations (PDC light curves). The differences in the light curves are negligible for the transit fitting in this work.

Our photometry pipeline starts by checking and correcting the stellar astrometry relative to the nominal position of the host star of the exoplanet as measured by Gaia. Our photometry measurement is then taken from a circular aperture of 3 pixels, corresponding to 63$\arcsec$. The sky background is accounted for as the median value of the pixels which constitute the lowest fifth percentile in flux in the vicinity of the host star. The quadrature sum of the standard deviation of these pixels and the Poisson noise of the source itself is used as the photometric uncertainty.

Contamination flux from nearby unresolved stars is removed based on a relationship between the flux brightness profile and the distance to the Gaia centroid of the unresolved stars. The flux percentage removed is 31.49$\%$$\pm$1.79$\%$, 0.2$\%$$\pm$0.007$\%$ for WASP-121b, and HD-219666b respectively. Possible blending from a binary companion should be subtracted as well. However, for these two sources, there is no evidence of a binary companion in earlier literature \citep{Delrez,Esposito2019}.

\begin{figure}
\includegraphics[width=3.5in]{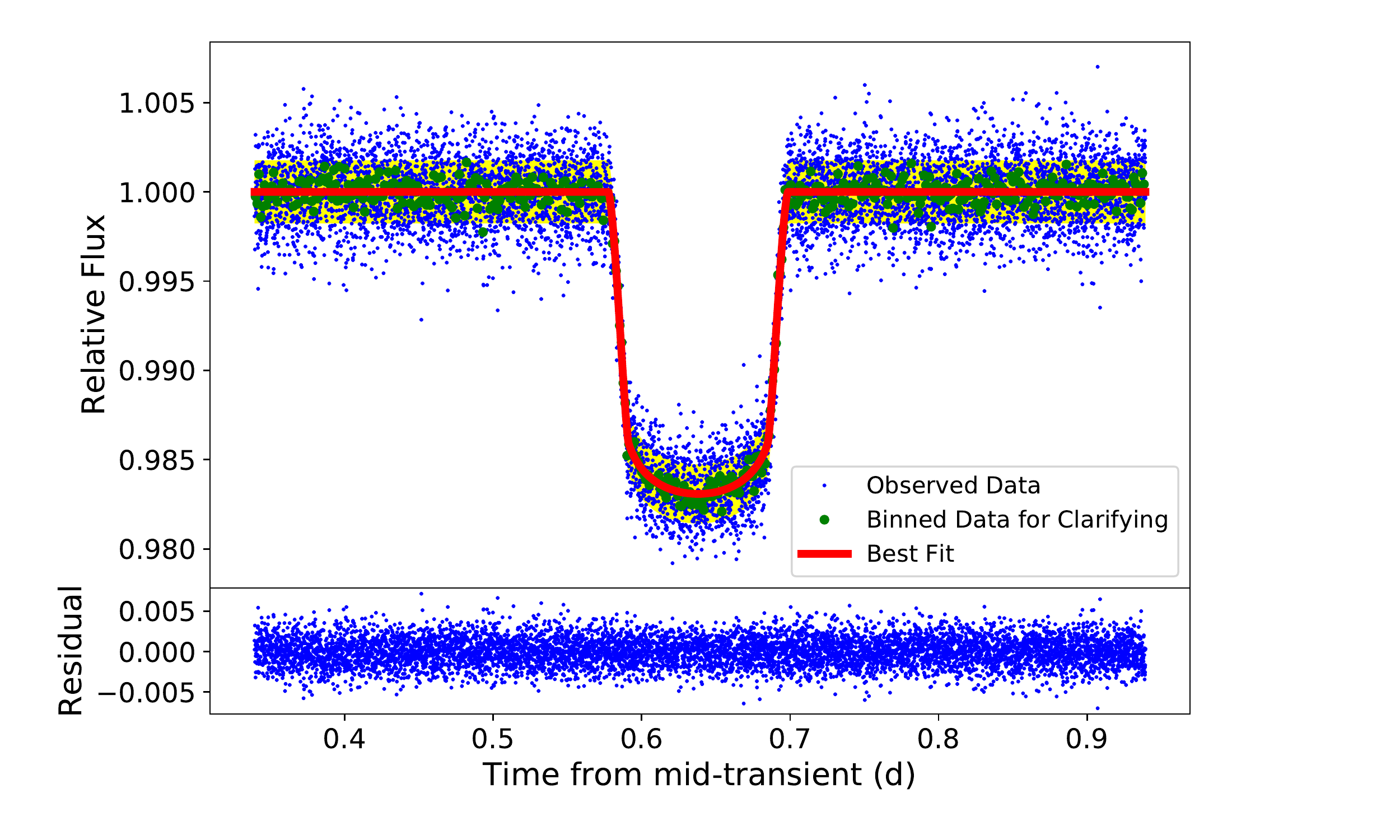}
\includegraphics[width=3.5in]{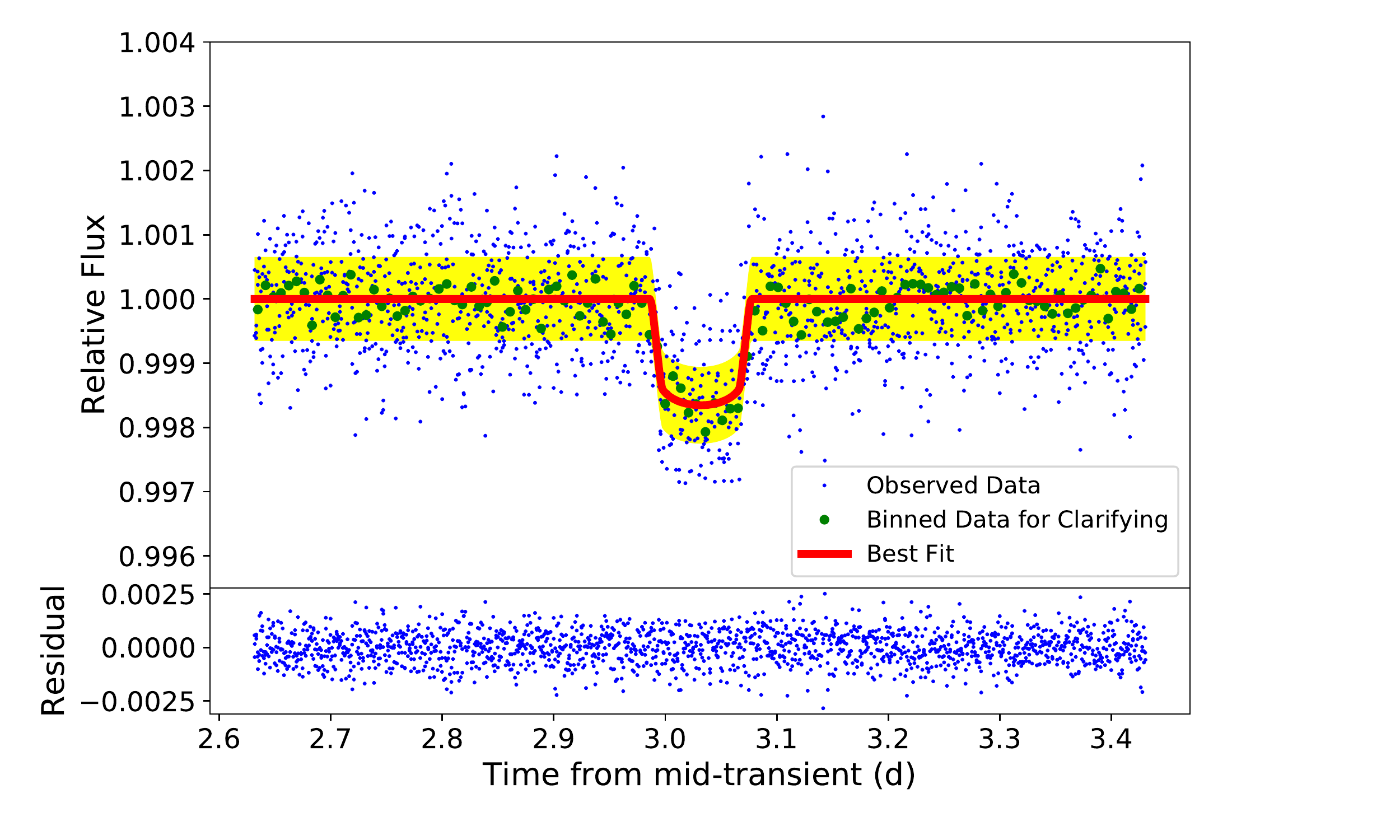}
\caption{Fits to the TESS transit folded light curves of WASP-121b (top), HD-219666b (bottom). The lower panel in each plot gives the residual map. The blue points are observed data; the green points binned to 30 minutes cadence for clarifying; the red solid line, the best-fit transit model; the yellow region, 1 $\sigma$ confidence region of the fits. The midpoint of the transit event is shifted to half the period, at the center of the x-axis.}
\label{LC}
\end{figure} 

Detrending the light curve to remove the long-term trends is needed for exoplanet transit fitting \citep{GPdetrending, SFF}. 
WASP-161b has well measured ephemeris for transits. This enables us to apply a simple but effective detrending algorithm based on ephemeris \citep{Yang2020}.
%Our targets are identified exoplanets that enable us to apply a simple but effective detrending algorithm based on the transit time stamps from ephemeris \citep{Yang2020}. 
We extract the light curve centered at the transit mid-point time. The light curves during the transit are then masked and fitted with a linear function. The fitted function is used to apply a correction to the light curve including the transit phase. Then, we fold the light curve around the transit phase, as shown in Figure \ref{LC}. We also perform tests with higher-order polynomial functions (up to order 10) and a cubic spline function \citep{Daylan2019} which give negligible differences when fitting the transit light curves.

\subsection{The application of iterative method to TESS light curves}

We apply our new method to TESS light curves of WASP-121b (transit depth SNR $\sim$ 9.6). We present HD-219666b (transit depth SNR $\sim$ 2.7) using non-iterative transit fitting as a comparison. The systems have already been identified as containing exoplanets, and so we take the same period, orbit type (circle orbit) and binary companion contamination from the discovery papers \citep{Delrez,Esposito2019}. 

The difference between the iterative method to the non-iterative method is as described in Section 2.1. The operation in a single iteration is the same as the classic MCMC fitting. A single iteration requires 50,000 steps, ignoring the first 30,000 steps as burn-in. The number of steps is determined by a trial run which shows that the chain is stable after the first 30,000 steps. The trial also indicates a negligible difference if we run for yet more steps, e.g. 200,000. The free parameters in MCMC fitting are $R_{p}/R_{\ast}$, the inclination of the planet orbit (i), the semi-major axis in stellar radii ($a/R_{\ast}$), time of transit center (T0, the transiting midpoint is shifted to half of the period) and the limb darkening parameters. The prior for the free parameters except limb darkening are all uniform (shown in Table 1).

%Specifically, the prior of the inclination is cut off at 90 degree which is commonly preformed in literature. We make test to change the prior to be $\mathcal{U}$[70,110]. The changes enlarges the uncertainty of posterior and induce a negligible bias with the prior $\mathcal{U}$[70,90]. The symmetry of inclination triggers a one-side prior in transit fitting to avoid two modality .  

An inclination prior of $\mathcal{U}$[70,90] is commonly used in the literature but is changed to be $\mathcal{U}$[70,110] for WASP-121b. The prior cut-off at an inclination of 90 degrees is to avoid the posterior presenting two symmetry modalities centered at 90 degrees. When the real inclination is close to 90 degrees, the posterior distribution is severely affected by the cut-off. Also, the two modalities are both very close to 90 degrees and negligible. The WASP-121b inclination shows a more reasonable posterior distribution without a 90-degrees cut-off in prior.

\begin{table*}
\setlength{\tabcolsep}{3mm}
\begin{center}
\caption{Free Parameters of Light Curves Fitting}
\label{tab1}
\begin{tabular}{ccccc}
  \hline
  \hline
\small
{Parameters }&       {Description } &      WASP-121b & WASP-121b & HD-219666b  \\
             &                      &  
Iterative Method  & Non-iterative & Non-iterative   \\ 
\hline
  \multicolumn{5}{c} {Parameter Prior} \\
  \hline
$R_{p}$/$R_{\ast}$ & Planet/star radius ratio& $\mathcal{U}$[0,0.2]      & $\mathcal{U}$[0,0.2]      & $\mathcal{U}$[0,0.1]  \\  
i   & Inclination   & $\mathcal{U}$[70,110] & $\mathcal{U}$[70,110] & $\mathcal{U}$[70,90]\\
a/$R_{\ast}$       & Semi-major axis of planet orbit in stellar radii & $\mathcal{U}$[0,30]        & $\mathcal{U}$[0,30]        &  $\mathcal{U}$[13.3,0.3] \\ 
T0 &  time offset of transit center & $\mathcal{U}$[0.3*1.27,0.6*1.27]      &  $\mathcal{U}$[0.3*1.27,0.6*1.27]      &   $\mathcal{U}$[0.3*6,0.6*6] \\ 
\textit{u}$_{1}$ & Linear limb-darkening & ...&$\mathcal{N}$(0.33,0.05) &$\mathcal{N}$(0.33,0.05)\\
\textit{u}$_{2}$ & Quadratic limb-darkening &... & $\mathcal{N}$(0.21,0.05) &$\mathcal{N}$(0.20,0.05)\\
\textit{c}$_{1}$  & Quadratic limb-darkening coefficient & ...& $\mathcal{N}$(2.84,0.05)&...\\
\textit{c}$_{2}$  & Quadratic limb-darkening coefficient & ...&$\mathcal{N}$(-4.93,0.05) &...\\
\textit{c}$_{3}$  & Quadratic limb-darkening coefficient & ...& $\mathcal{N}$(4.77,0.05)&...\\
\textit{c}$_{4}$  & Quadratic limb-darkening coefficient & ... & $\mathcal{N}$(-1.64,0.05)&...\\  
\hline
  \multicolumn{5}{c} {Fitting Result (quadratic law)} \\
  \hline
\textit{u}$_{1}$  & Linear limb-darkening coefficient & 0.24$\pm$0.03 & 0.23$\pm$0.04 &0.32$\pm$0.03\\
\textit{u}$_{2}$  & Quadratic limb-darkening coefficient &  0.09$\pm$0.03& 0.16$\pm$0.04 &0.19$\pm$0.03\\

$R_{p}$/$R_{\ast}$ & Planet/star radius ratio& 0.1239$\pm$0.0003      & 0.1234$\pm$0.0004 &0.0418$\pm$0.0004  \\  
i   & Inclination   & 89.5$\pm$1.5 & 89.9$\pm1.6$ & 86.45$\pm$0.13\\
a/$R_{\ast}$       & Semi-major axis of planet orbit in stellar radii & 3.80$\pm$0.03   &  3.81$\pm0.03$   &  13.33$\pm$0.31 \\ 
std of residual &  Parts per million  &  1763 &  1763 & 652 \\ 
reduced $ \chi^{2}$ &  Reduced chi-square  &  1.0004 &  1.0004 & 1.0005 \\
\hline
  \multicolumn{5}{c} {Fitting Result (non-linear law) $^{a}$} \\
  \hline
 \textit{c}$_{1}$  & Quadratic limb-darkening coefficient & 1.54$\pm$0.04& 2.81$\pm$0.04 &...\\
\textit{c}$_{2}$  & Quadratic limb-darkening coefficient & -2.42$\pm$0.04& -4.96$\pm$0.04&...\\
\textit{c}$_{3}$  & Quadratic limb-darkening coefficient & 2.28$\pm$0.04& 4.74$\pm$0.04&...\\
\textit{c}$_{4}$  & Quadratic limb-darkening coefficient & -0.76$\pm$0.04 &  -1.67$\pm$0.04&...\\ 
$R_{p}$/$R_{\ast}$ & Planet/star radius ratio& 0.1237$\pm$0.0003      & 0.1234$\pm$0.0003 &...  \\  
i   & Inclination   & 89.9$\pm$2.0 & 90.0$\pm1.8$ & ...\\
a/$R_{\ast}$       & Semi-major axis of planet orbit in stellar radii & 3.80$\pm$0.03   &  3.81$\pm0.03$   &  ... \\ 
std of residual &  Parts per million  &  1763 &  1763 & ... \\ 
reduced $ \chi^{2}$ &  Reduced chi-square  &  1.0004 &  1.0004 & ... \\
\hline
\end{tabular}
\end{center}
\begin{center}  
a: Quadratic law only fitted for HD-219666b, to be consistent with identification paper \citep{Esposito2019}. 
\end{center}
\end{table*}

\subsection{Convergence, multimodality for quadratic and non-linear models}

We utilise the iterative method with both the quadratic and no-linear LD laws. These laws are expressed by the following equations.
\begin{equation}
\frac{\mathit{I}(\mu)}{\mathit{I}(1)} = 1 - \mathit{u_{1}}(1-\mu)-\mathit{u_{2}}(1-\mu)^{2}
\textup{ \ \ \ (quadratic law),}
\end{equation}
\begin{equation}
\frac{\mathit{}{I}(\mu)}{\mathit{}{I}(1)} = 1 - \sum_{n=1}^{4}c_{n}(1-\mu^{n/2})
\textup{ \ \ \ \ \ \ \ \ \ \ \ \  (non-linear law),}
\end{equation}
where $\mu$ is give by cos($\gamma$), $\gamma$ is the angle from the outward surface to our line of sight, $u_{1}$, $u_{2}$ and $c_{n}$ are the LDCs. 

The iteration is taken as convergent if the transit parameters reach a certain set of values and then only show small fluctuations subsequently. In this work, the convergence is treated as achieved if the fluctuation from loops 25 to 30 of the iteration is less than 0.005 (calculated from the LDC prior's $\sigma/10$). The iteration is convergent for WASP-121b when modeled with the quadratic law (as shown in Figure \ref{itpro}). The LDCs are constrained by the transit light curve to some extent when applying the non-iterative method \citep{muller2013, Espinoza2015, Evans2018}. Otherwise, the transit fitting in earlier works makes non-sense. The partial constrain to LDCs in every single iteration fitting can be enlarged with our iterative method and thereby forms an LDC prior independent fitting method.

For the non-linear law, the transit parameters do not change much during an iteration, within 0.5$\sigma$ derived from MCMC fitting. The transit parameters except the LDCs reach convergence in the first loop. This is consistent with the conclusion from earlier literature \citep{Espinoza2015} that a more complex LD model gives more reliable transit parameters for non-iterative (single iteration) method. 

However, the LDCs for the non-linear law do not converge to specific values. The standard deviation (std) of the fitting residual is about the same 
for different LDCs, implying multimodality which is a typical issue in statistics \citep{Geyer}.

The multimodality of LDCs does not appear in a single MCMC fitting. MCMC having difficulty in jumping between the local modes is a common MCMC occurrence \citep{mcmcmultimodality, Metropolis}. This increases the risk of underestimating the uncertainty in the LDCs and other parameters. How to resolve this is a much-debated topic in the field of machine learning \citep{tempering2006,mcmcmultimodality,paulin2019}. 
%A comprehensive analysis of the non-linear law model needs Monte-Carlos simulation which is described in Sec.3. 

%Among our fitting, the MCMC retrieve LD with $\sigma$$\sim$0.03 in any ititial input prior with a $\sigma$=0.05. However when we repeated the fitting for 100 times, the standard deviation of the same LD is 1.5, 5 times larger than the $\sigma$ from MCMC which indicates that the chain is stuck in one mode during MCMC fitting. Also, We tried 5 different group initial LD input, the iteration took separate LD paths, but all retrieve same radius ratio, inclination and semi-major axis after a few loops. The result implies that the MCMC fitting with a prior of $\sigma$=0.05 obtain sets of light curve preferred LD and other transit parameters.  

%We change the LD prior $\sigma$ to be 0.01 during the iteration to remove the different modes during MCMC fitting. Handling with multimodality is a on-going field in modern statistics \citep{tempering2006,mcmcmultimodality,paulin2019}, which is beyond the purpose of this work. We aim to find the light curve preferred LD to derive the unbiased $R_{p}/R_{\ast}$, inclination, and semi-major axis. As decribed in the obove paragraph, one of the precise degenerate LD modes is good enough to this end. Better than that, when we reduce the prior boundary of the fitting,  

\subsection{Limb darkening result}

The estimates of the transit parameters are taken from the "final fitting" (as shown in Table 1). To arrive at our final transit fitting, for the quadratic law we take the LDC values at convergence, while for the non-linear law we take the LDC values after the 5th iteration. The fitted light curve when applying the quadratic law is as shown in Figure \ref{LC}. The fitted light curve using the non-linear law looks very similar and so is not included.

\begin{figure}
  \centering
        \includegraphics[width=3.5in]{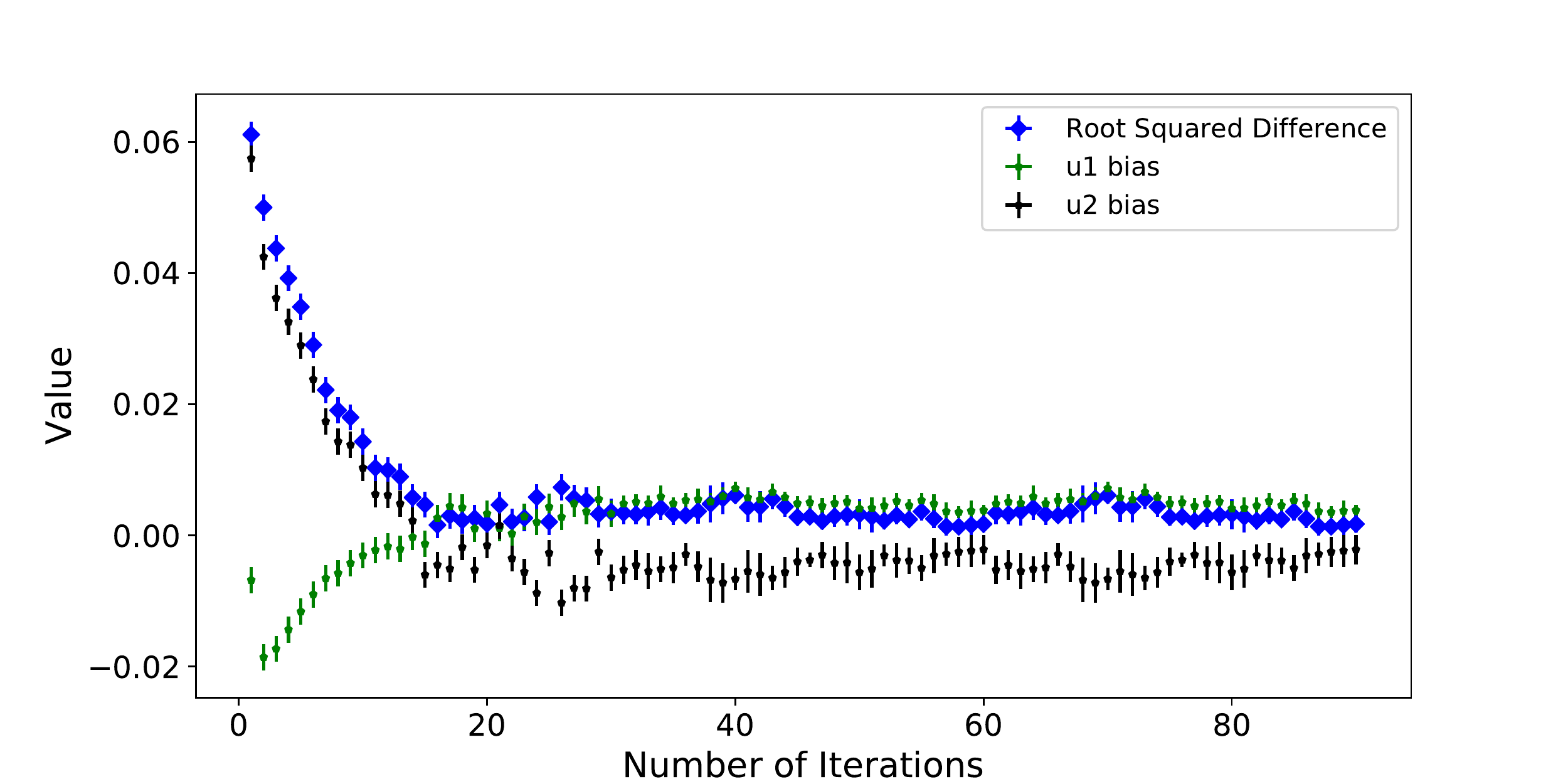}
 \caption{The iteration process for the quadratic law. The blue diamonds show the root squared difference (as defined by equation \ref{equ:rds}); the green points, u1; black points, u2. The error bar of each coefficient is the standard deviation of the parameter from 10 fittings with the same input prior. The error bar of root squared difference is derived from the error of each LDC following the error propagation.}
 \label{itpro}
\end{figure}

Using the convergence definition in the previous section, the quadratic law converges. To assess the speed and convergence of the iterative method, we define root squared difference as:
\begin{equation}
Root \ Squared \ Difference = \sqrt{\sum (a_{i,n}-a_{i})^{2}}   
\label{equ:rds}
\end{equation}
a$_{i,n}$ is LDC$_{i}$ at the nth iteration loop, a$_{i}$ is the median value of LDC$_{i}$ when the iteration is converges. The iteration procedure is shown in Figure \ref{itpro}. Each parameter shows an individual path to convergence. The iterations when converging show a fluctuation within 0.005.

%\textit{The lost function in each loop with same prior has a $\sigma_{c}$ $\sim$ of 0.013 for the quadratic law and 0.02 for the non-linear law. The MCMC fit LD derives a parameter $\sigma$$\sim$0.042 for both laws. From error propagation, we know the median value of a parameter should have a factor of square root of data number smaller uncertainty. So far, the assessment of the LD estimation are, but only self-consistent internal our method. A further evaluation of the LD value and uncertainty is performed, using a Monte Carlo simulation in Section 3. }

This work aims at discussing and reducing the bias of transit parameters to within a few $\sigma$. It is necessary to prove that our fitting does not induce extra uncertainties compared to the published works. To this end, HD-219666b is used for comparison purposes. We fit the light curve with the non-iterative method using the same prior as the identification paper \citep{Esposito2019} but replacing the median value of the prior with their reported result. This exercise is to understand whether we can obtain the same fitting parameter estimates when applying the same data and fitting method. The fitting applies the quadratic law, as the same as \citet{Esposito2019}. The transit parameter estimates from our MCMC fitting, as shown in Table 1, are consistent with \citet{Esposito2019} to within 0.3$\sigma$. We try the iterative method on HD-219666b but find it does not converge. This non-convergence is mostly due to the low SNR. A comprehensive analysis of convergence using simulations is described in the next section.

We also show the result for WASP-121b following the classic transit fitting method (no iteration), applying both the quadratic and non-linear laws. The LDC priors use median values, taken from the atmospheric models of the host star \citep{TESSLD}, with a Gaussian distribution $\sigma$ of 0.05. From the results (as shown in Table 1), we find that transit parameters' posterior values are highly consistent (within 0.1$\sigma$) with the quadratic and non-linear laws when applying the LDCs from the same stellar atmospheric model. 

The results from the iterative and non-iterative methods show significant differences in the LDC posteriors. More importantly, the difference in $R_{p}$/$R_{\ast}$ between iterative and non-iterative methods is 0.4$\%$, 1$\sigma$ for the quadratic law; and 0.2$\%$, 0.6$\sigma$ for the non-linear law. This difference level is crucial to take into considering when analyzing the exoplanet atmosphere \citep[Yang et al. 2019, submitted; ][]{seager2010, Evans2018}. Simulations are used to help assess these differences (the next section).

\section{Assessment of LD applications Using Simulations}

Monte Carlo simulations are performed to evaluate the biases and uncertainties of LDCs and the other transit parameters, especially $R_{p}$/$R_{\ast}$, obtained by iterative and non-iterative methods. The LDC biases if present and error propagation to other parameters are most important to evaluate. The simulations generate light curves with certain inputs, mainly based on the fitting result of WASP-121b (as shown in Table 1). Gaussian random noise is added to the light curve.

\subsection{LDC measurement from iterative method}

From our experiences on WASP-121b with real data, the LD quadratic law converges but the non-linear law does not. We evaluate the LDC measurement using Monte Carlo simulations.

Simulated light curves are constructed by applying \citet{Mandel_Agol2002} models with derived transit parameters of WASP-121b obtained with the iterative method (as shown in Table 1). Time sampling is the same as TESS: 2 seconds sampling but binned to 2 minutes. Random noise with a $\sigma$ of 1763 parts per million (ppm) is added to the light curve. We apply the quadratic law iterative method as described previously. We repeatedly generate and fit new light curves 100 times in total. The starting LDC priors are randomly chosen and different every time. The MCMC process takes a few days to complete running on an 8-core processor with a 2GHz clock frequency.

\begin{figure}
  \centering
        \includegraphics[width=3.5in]{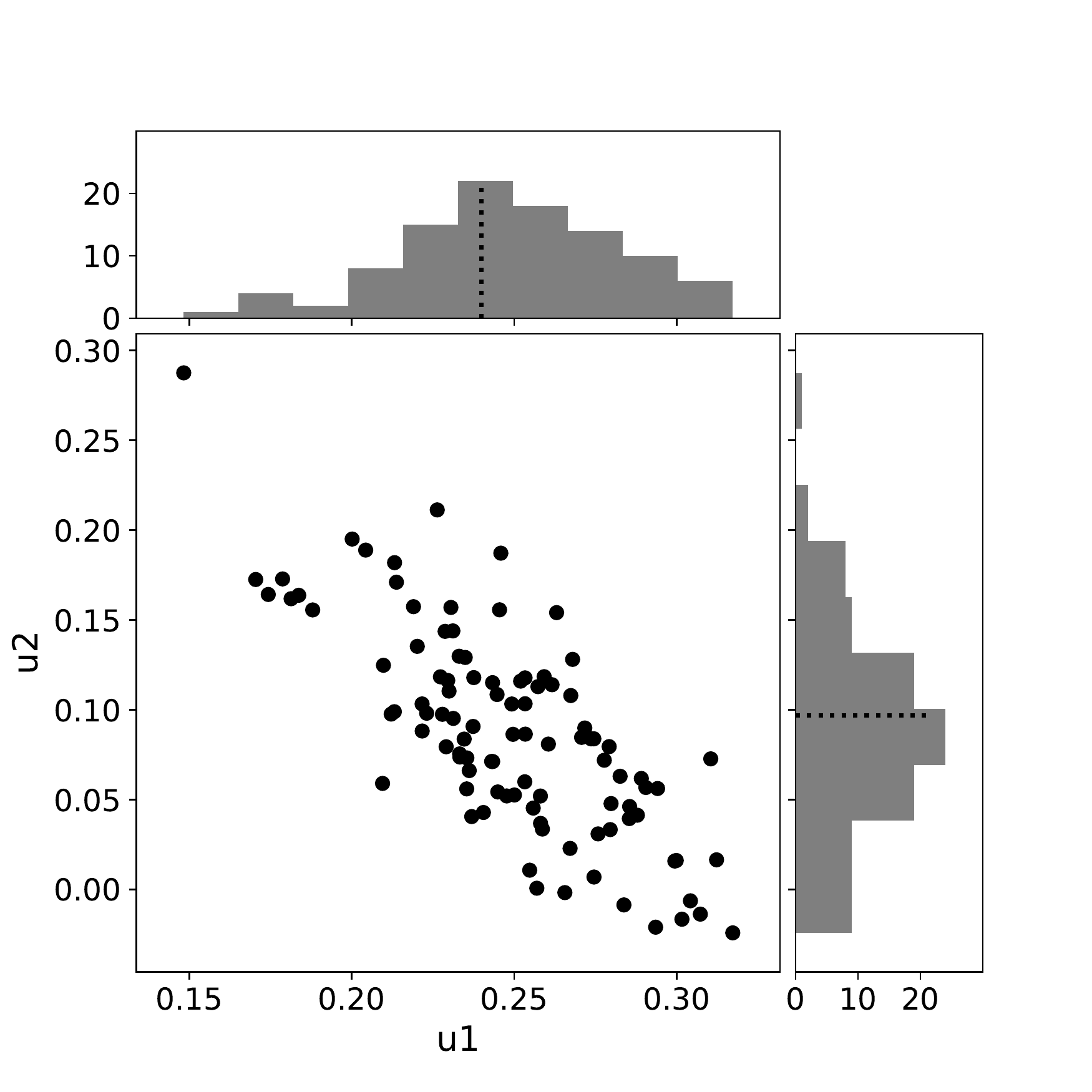}
 \caption{The distribution of LDCs from 100 simulations. The dash lines in each histogram indicate the input value of the LDCs: 0.24 for \textit{u}$_{1}$, 0.09 for \textit{u}$_{2}$. The estimated values of \textit{u}$_{1}$, \textit{u}$_{2}$ are 0.25$\pm$0.03, 0.08$\pm$0.06, respectively.}
\label{LDCdis}
\end{figure}

Out of the 100 modeling runs, 99 converge. Convergence is taken as achieved if the LDC fluctuations between interation loops 25 to 30 is less than 0.005. The distribution of the LDCs when converged is shown in Figure \ref{LDCdis}. The LDC values are for 0.25$\pm$0.03 for \textit{u}$_{1}$, 0.08$\pm$0.06 for \textit{u}$_{2}$ which are very close to the input LDC values as 0.24 for \textit{u}$_{1}$ and 0.09 for \textit{u}$_{2}$. 
%Also the $\sigma$ of LD distribution is consistent with the uncertainty estimation from the final fitting to the real data applying the convergent LD as priors. The result implies that applying a $\sigma$ of 0.05 for LD prior is reasonable.  

As with real data, the non-linear law applied iteratively does not converge for the simulated light curves. The LDC input values are set as the result from the real light curve modeled with the iterative non-linear law. The input values of other parameters are taken from the quadratic law result. The analysis in this subsection from now on refers to quadratic law only.

We thus conclude that the quadratic law iterative method arrives at a correct LDC estimation, at least for WASP-121b TESS data. In the simulation test, we cover the parameters we see as important. The effectiveness of our iterative method depends on photometry accuracy, binning, and the transit parameters, e.g. transit duration, inclination, stellar surface brightness (described by LDCs). Empirically the parameters above are the most crucial in our topic, although more parameters affect the transit fitting. Photometry accuracy includes two parts, i.e. the noise in the photometry and the number of measurement points. In statistics, the two parts are connected by the relation
\begin{equation}
\sigma_{l}=\sigma_{s}/\sqrt{n}, 
\label{error propagation}
\end{equation}
where $\sigma_{l}$ is the noise of data sets with larger binning interval; $\sigma_{s}$, the noise of data sets with smaller binning interval; \textit{n}, the ratio of the amounts of data sets. In the simulations, we take the same number of data points as the real data of WASP-121b. The light curves with worse photometry accuracy result in lower SNR of $R_{p}$/$R_{\ast}$.

We change the input values of these parameters in simulations, one at a time, and fit the light curve to see if we obtain output consistent with the input. For each set of parameters, we repeat the simulation 10 times to ensure statistical significance. 

Our fitting results indicate that the output LDCs are consistent with the input values except for light curves with low SNR, light curves with high impact parameter ($b$), and light curves with unusually large LDC values. 

The quadratic law iterative method does not converge when SNR below 5. The light curves have insufficient SNR and/or a number of data points to model the limb darkening properly when SNR $\sim$5. The std of the residual of the fitted light curve is the same as the noise added in the light curve. The well fitted light curves with different sets of LDCs hint at LDC multimodality. The simulations imply that 8 is the lower limit for SNR to ensure that the iterative method converges when the transit parameters are the same as WASP-121b.

The morphological distortion of light curves caused by finite integration times has been considered by \citet{David2010}. We have discussed inclination and semi-major axis biases versus the binning in the TESS WASP-121b light curve in earlier work (Yang et al. 2019, submitted). We concluded that the 30 minutes cadence caused significant bias to the estimation of inclination and the semi-major axis for WASP-121b. The binning threshold for non-bias detection in the inclination (semi-major axis) strongly depends on the transit duration which is 5 minutes for WASP-121b.

\begin{figure}
  \centering
   \includegraphics[width=3.5in]{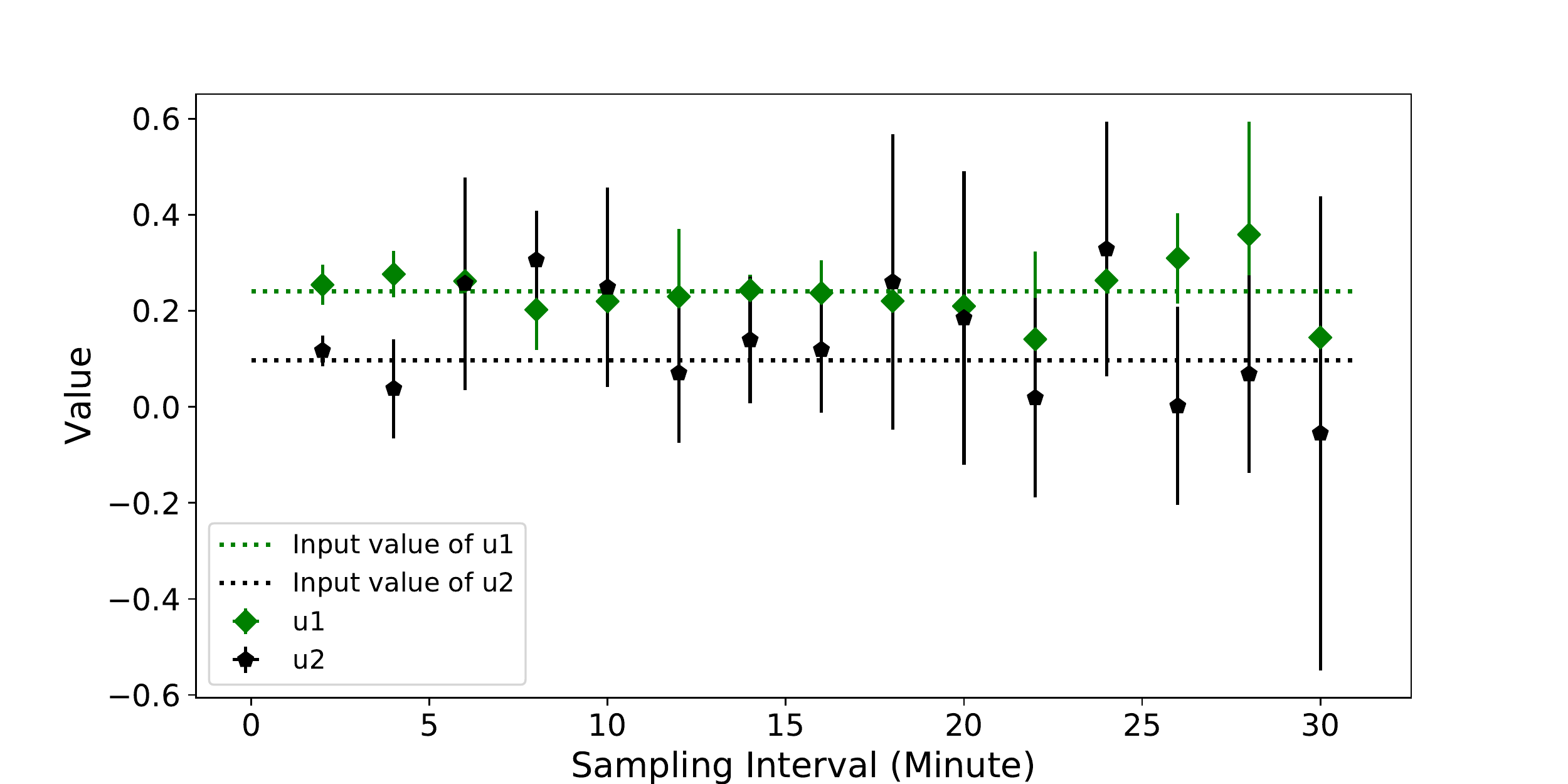}
 \caption{The LDC distributions for different binning intervals. The simulation of light curves including generating and fitting is repeated 10 times for each binning interval. The LDCs are obtained as the median values among 10 times repeated simulations. The error bars are obtained as LDC standard deviations among 10 simulations.}
\label{LD_sampling}
\end{figure}

We have assessed the impact on the iterative method of simulated light curves with a binning interval of fewer than 5 minutes. The iterations do converge and parameters are obtained consistently. For the simulations with a binning interval longer than 5 minutes, we find that the iterations converge, and the LDCs obtained do not show any significant bias to the input value as shown in Figure \ref{LD_sampling}. 

A high impact parameter causes difficulties in transit fitting because the transit happens too close to the limb of the star. The impact parameter is defined as :
\begin{equation}
    b= \frac{ a \times cos(i) } {R_{\ast}},
\end{equation}
where $a$ is semi-major axis, $i$ refers to inclination, $R_{\ast}$ indicates stellar radius. The LDCs from light curve fitting are significantly different from the stellar model predicted LDCs when the impact parameter is high \citep{muller2013, Espinoza2015}. In our simulations, the iterative method fails to converge when the inclination is close to 75 degrees which is the highest impact parameter among all possible transits. The situation is straightforward to understand when the transit just grazes the edge (limb) of the host star: the stellar surface provides little useful input to the transit modeling. Uncertainty in the LDC values propagates severely to the other transit parameters, especially $R_{p}$/$R_{\ast}$, which disrupts the whole modeling process.

%We find that the iteration is convergent and reaches a biased LD when the inclination is 80 equivalent to an impact parameter of 0.66. The uncertainty of the parameters is larger than when the inclination is 89.5. The $R_{p}$/$R_{\ast}$ yields an uncertainty of 0.0013, 4 times larger than the result when inclination being 89.5. The converged LD are 0.40$\pm$0.05 for \textit{u}$_{1}$ and 0.08$\pm$0.10 for \textit{u}$_{2}$, according to the distribution of iteration results of 10 simulation light curves. 

The iteration mechanism works for general cases of LD but becomes less effective when both linear and quadratic coefficients are larger than 0.4. Such LDCs would be appropriate for log \emph{g} $\sim$ 2.5 cgs, and T$_{eff}$ $\sim$ 3000 K \citep{TESSLD} which are uncommon, and very different from the main sequence stars \citep{GaiaHR, Zhang2020}. 
%\citet{Howarth2011} claimed that the limb darkening coefficients obtained from photometry light curves and from model stellar atmospheres are not directly comparable because of different optimization processes. 
%The intrinsic difference as a result of the two optimizations is beyond the scope of this work. For scientific purposes, the LD application which yields better transit parameter estimates is the one we should utilize. 

%Essentially, it is more important to derive precise transit parameters especially $R_{p}$/$R_{\ast}$ with a proper application of LD. A research on the influence of different LD application towards other transit parameters is described in the following subsection.  

\subsection{iterative method influence on transit parameters}

Ideally, transit modeling, including limb darkening modeling, should provide unbiased estimates of all parameters, and most importantly $R_{p}/R_{\ast}$. We perform simulations to assess the bias of transit parameters.

\begin{figure}
  \centering
   \includegraphics[width=3.5in]{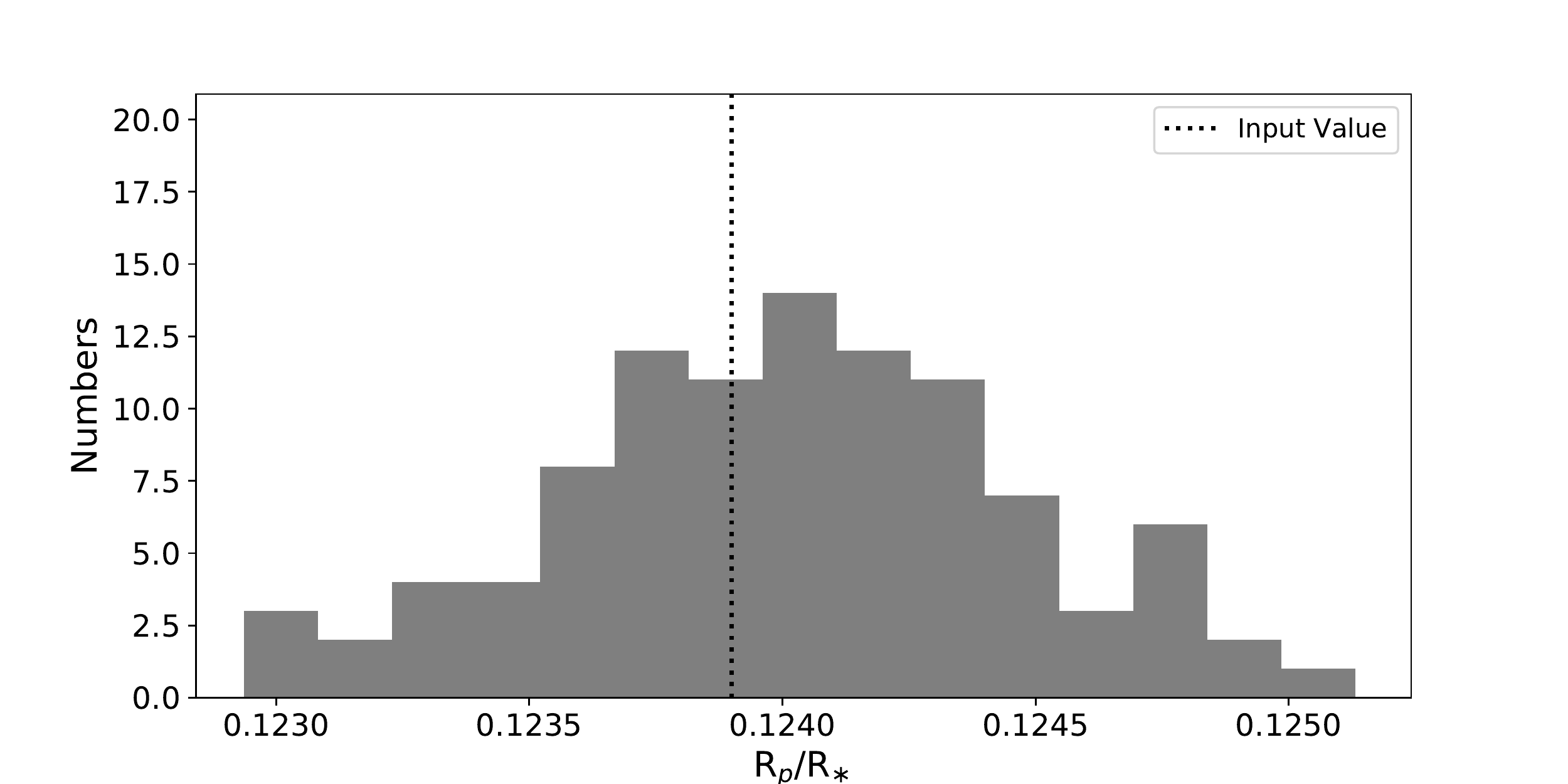}
 \caption{The $R_{p}/R_{\ast}$ distribution from using the iterative method with 100 simulated light curves. The vertical line indicates the input value of $R_{p}/R_{\ast}$ at 0.1239.}
 \label{radiusdis}
\end{figure}

The simulations are performed using the quadratic law iterative method to check if the method can derive unbiased transit parameters. The simulation is the same as the one which gives the LDC distributions in Figure \ref{LDCdis}. We generate light curves and fit them with arbitrary initial LDC priors, repeating the simulations 100 times. The estimated transit parameters are consistent within 0.66$\sigma$ with the input values to construct the light curves. The $R_{p}/R_{\ast}$ appears at 0.1240$\pm$0.0004 which differs at 0.25$\sigma$ to the input value (as shown in Fig. \ref{radiusdis}). The differences in inclination and semi-major axis are at 0.35$\sigma$ and 0.66$\sigma$. The simulation shows that the iterative method is effective in estimating unbiased transit parameters.

Moreover, the transit parameters (except LDCs) from the iterations after the first few are close to the final converged result. The difference in $R_{p}/R_{\ast}$ between the 5th, 10th, 15th and final $R_{p}/R_{\ast}$ is negligible at 0.2$\sigma$ level. The result indicates that the LDCs after 5 loops as well as the converged LDCs are all suitable for citing in the "final fitting" if aiming at unbiased transit parameters like $R_{p}/R_{\ast}$. 

The transit parameters are not biased ($R_{p}/R_{\ast}$ within 0.3 $\sigma$) if the iteration converges with changed photometry precision, transit duration, inclination, and LDCs when constructing the simulated light curves. The simulations also reveal that the transit parameters are usually not biased even if the LDCs do not converge. The bias of transit parameters (except LDCs) arises when the inputs of simulated light curves are more extreme, e.g., SNR less than 1.25, each LDC larger than 0.4, impact parameter larger than 0.7. The light curve contains encoded information about an exoplanet's transit across the face of a star but it will not necessarily be able to constrain all transit parameters equally well. In particular, a light curve may not be suitable for estimating the limb darkening parameters but maybe still usable for estimating other transit parameters. For limb darkening, this unsuitability presents itself as non-convergence, possible multimodality. When suitability drops furtherly and the data is not really usable for other parameters, e.g. $R_{p}/R_{\ast}$, the estimates become biased.

Simulation tests with the non-linear law perform similarly in parameter estimating. The simulation is the same as described in Sec. 3.1. The transit parameters (except LDCs) are almost stable after the first iteration. We take the LDC values from the 5th, 10th, 15th iterations to be priors for the "final iteration". Unlike the unconverged LDC posterior parameters, the estimated $R_{p}/R_{\ast}$ is stable. The value is 0.1237$\pm$0.0004 which is 0.5$\sigma$ different from the input radius ratio. The differences in other parameters are within 1$\sigma$ compared to their values.

\subsection{Comparison with transit parameters obtained by non-iterative methods}

The simulation tests on the iterative method indicate that the transit parameter estimation is unbiased. Another important matter is whether or not the transit parameters estimated by non-iterative means are biased. Furthermore, if the bias is present, how large the bias is for each parameter, especially for $R_{p}/R_{\ast}$, needs to be understood.

A basic test is to take the values used in creating the light curves and use them as priors in a simulation and then check the posterior values after the simulations have completed. This enables us to establish whether the fitting itself is introducing any bias.

\begin{figure}
  \centering
        \includegraphics[width=3.5in]{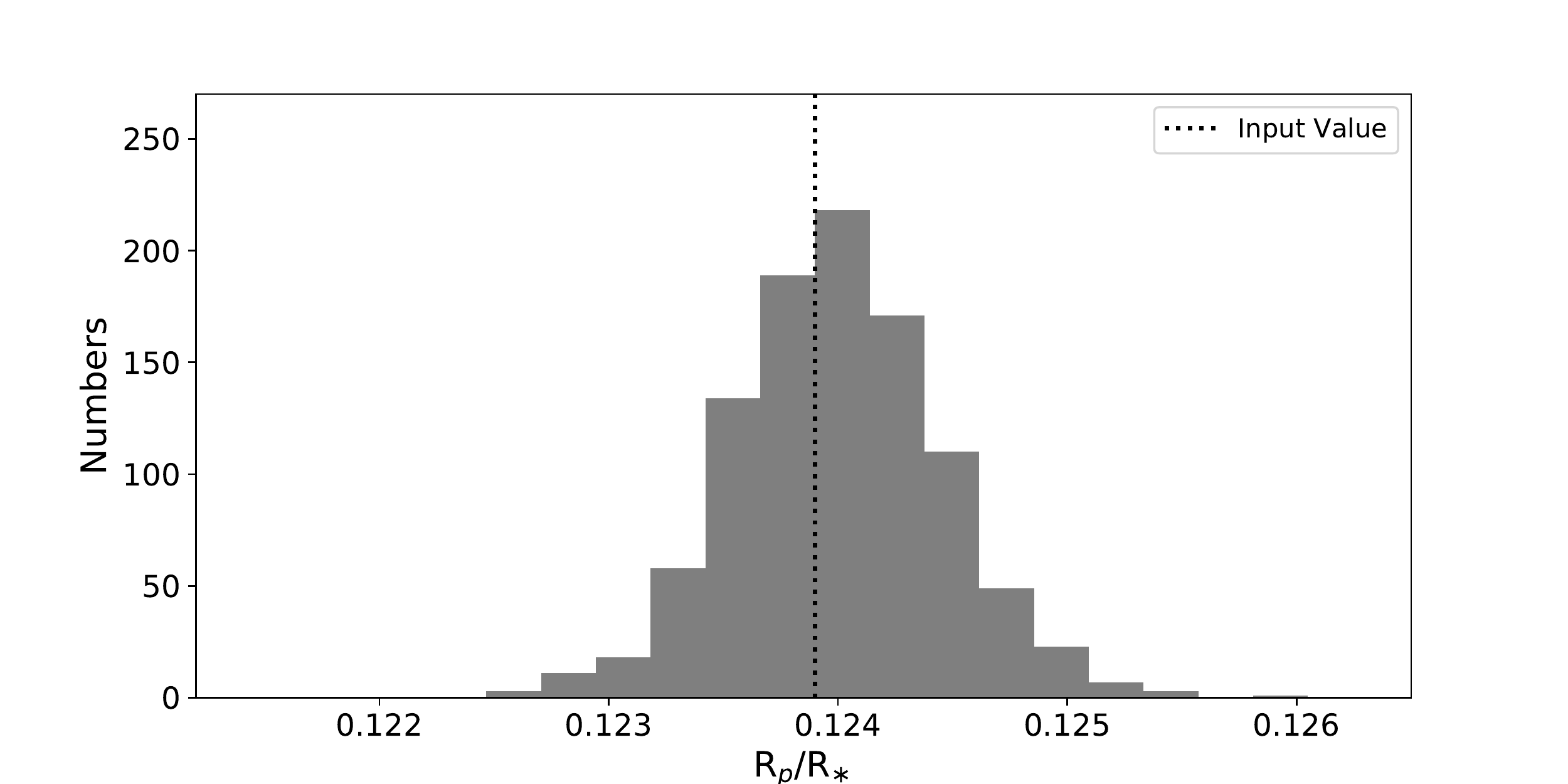}
 \caption{The $R_{p}/R_{\ast}$ distribution with non-iterative method for 1000 simulated light curves. The vertical line gives the input value.}
 \label{1000radius}
\end{figure}

We perform non-iterative fitting of 1000 simulated light curves following the procedure described in Sec. 3.1. The light curves are constructed with input values the same as the iteration result from using real data. The distribution of the results centers around the input value and forms a distribution due to the random noise added to the light curves and uncertainty in the fitting process. We show the distribution of $R_{p}$/$R_{\ast}$ as an example (in Figure \ref{1000radius}). It yields an $R_{p}$/$R_{\ast}$ value of 0.1240$\pm$0.0004 compared with the input value of 0.1239.

Also, the posteriors of the parameters in a single fitting just represent the distribution of transit parameters from the fitting to 1000 simulated light curves. The $\sigma$ of $R_{p}$/$R_{\ast}$ posterior is 0.0003 while the standard deviation of the $R_{p}$/$R_{\ast}$ derived from fitting to 1000 light curves is 0.0004. The $\sigma$ and standard deviation of each LDC are $\sim$ 0.03 and 0.03, respectively. The consistency between the LDC posterior from a single fitting and the statistical distribution of LDC values from multiple light curves demonstrates that the MCMC modeling is self-consistent.
%has been fed with a correct $\sigma$ for LDC prior.  

Similarly, we perform fitting for 100 simulated light curves generated using the non-linear law. The LDC priors have the same values as used in constructing the light curves. The distribution of $R_{p}$/$R_{\ast}$ centers at 0.1239, with an uncertainty of 0.0004.

The simulations with the non-iterative method show the fitting results centering at the input values if the priors are set to the same values as the input values. Fitting to the real data, we are not sure about the correctness of priors and the bias propagated to transit parameters. This impact needs to be evaluated through simulations. We, therefore, apply offsets to the input values and take the shifted values as the priors in simulations. We only apply offsets to the LDCs for the purposes of this work. The unshifted input values are 0.24 for \textit{u}$_{1}$, and 0.09 for \textit{u}$_{2}$.

\begin{figure}
  \centering
   \includegraphics[width=3.5in]{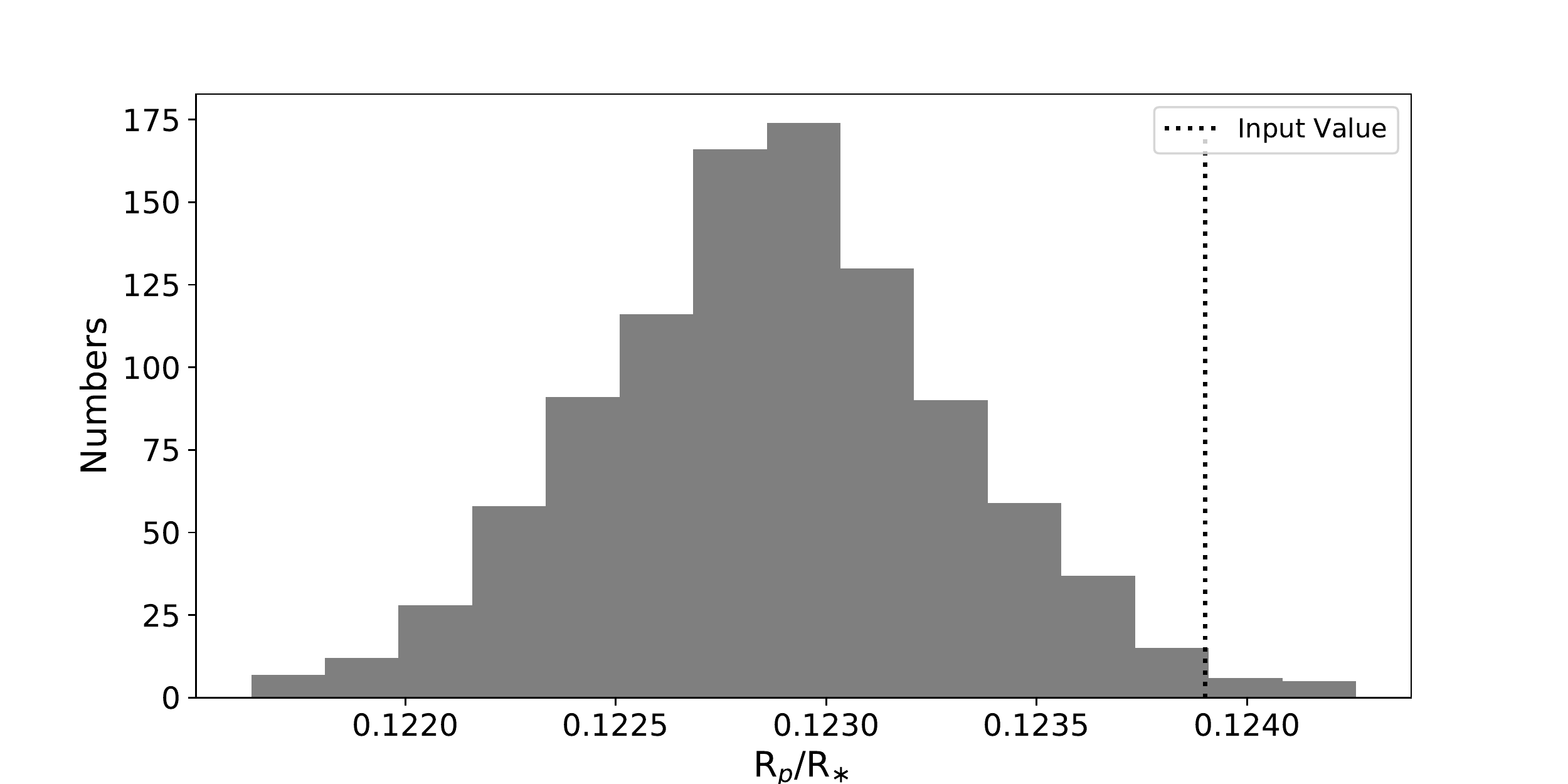}
        \includegraphics[width=3.5in]{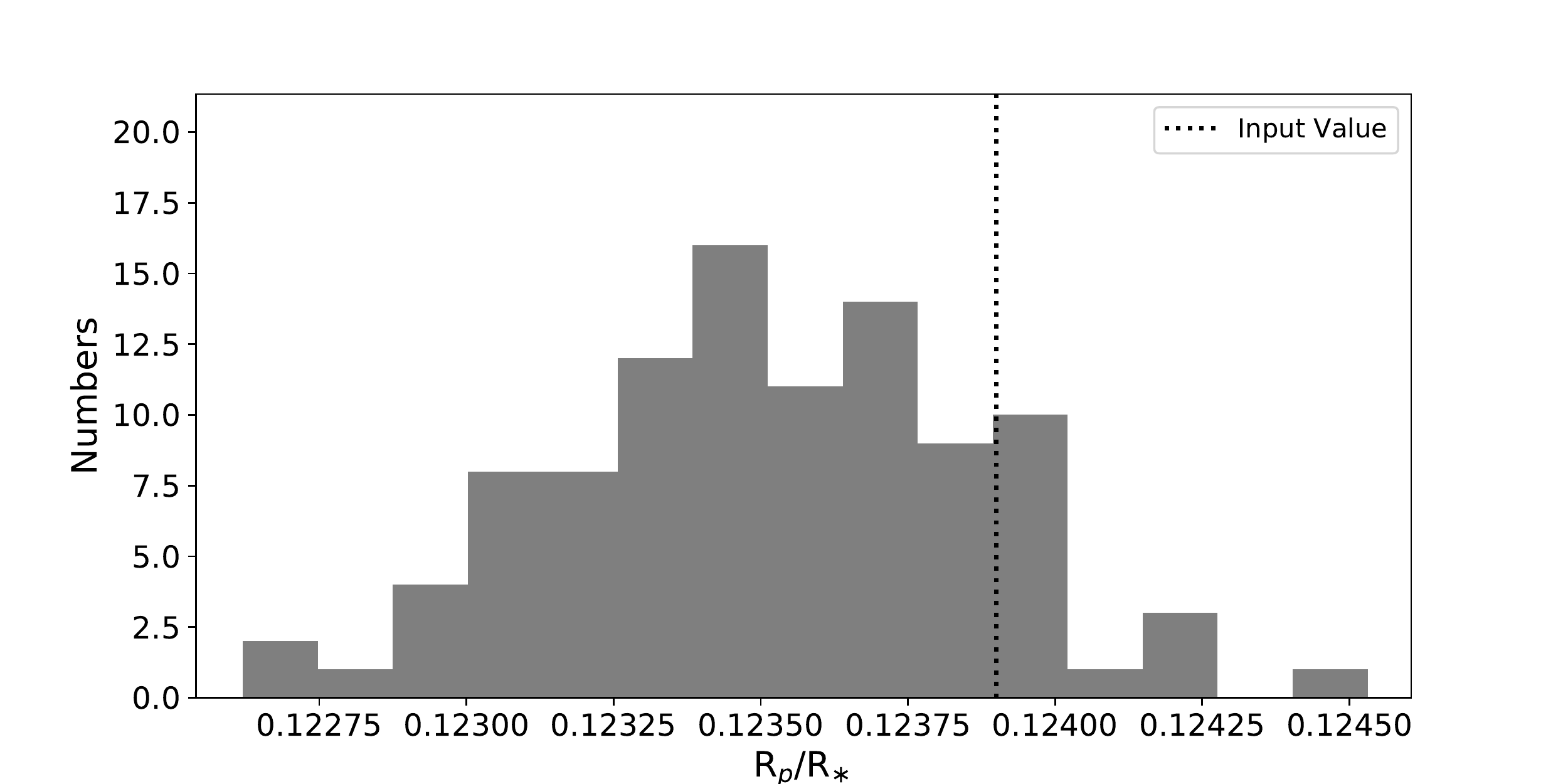}
 \caption{The $R_{p}/R_{\ast}$ distribution for fitting with shifted LDC prior: upper for quadratic law; lower for non-linear law. The vertical line gives the input value of $R_{p}/R_{\ast}$ at 0.1239.}
 \label{shiLD}
\end{figure}

For every offset of the prior, we generate and fit 100 light curves. The offset is with a step of 0.05 independently for \textit{u}$_{1}$ and \textit{u}$_{2}$. The fitting results imply that the bias in the fitted parameters is significant. The bias of $R_{p}/R_{\ast}$ is up to 0.82$\%$ which is at a significance level of 2.5$\sigma$ when the offset is 0.3 or larger for both \textit{u}$_{1}$ and \textit{u}$_{2}$ (as shown in Fig. \ref{shiLD}).

We emphasize a certain combination of priors, which is  
0.33$\pm$0.05 for \textit{u}$_{1}$, and 0.21$\pm$0.05 for \textit{u}$_{2}$. This set of LDC priors is just the same as the prior we applied in real data fitting with the non-iterative method. The simulation result implies a $R_{p}/R_{\ast}$ distribution centered at 0.1233 with a $\sigma$ of 0.0004. The results of fitting transit parameters are almost the same (within 0.25$\sigma$) as the fitting of real data. We thus conclude that the difference in $R_{p}/R_{\ast}$ between the iterative and non-iterative methods when fitting the real data is due to the offset of the LDC priors, and that $R_{p}/R_{\ast}$ is biased in non-iterative fitting at 0.5$\%$, 1.5$\sigma$.

We also generate and fit simulated light curves for 1000 times with a broad uniform prior, centered on the input LDCs and with an interval of 0.5. The parameters fluctuate too much and are not suitable for transit fitting. This implies that predictions of appropriate LDC priors are needed when applying non-iterative MCMC fitting to transit light curves, at least for the WASP-121b TESS light curve.   

In the case of the non-linear law, the simulation is performed similarly to the quadratic law. The result shows that the offset of the LDC priors biases the obtained $R_{p}/R_{\ast}$ but not as much as in the quadratic law. The distributions of radius ratios among simulations with different LDC priors are about the same. We take one set of biased LDC prior as an example. The LDC prior is shifted with an offset of 0.2 for each coefficient. Generating and fitting the light curve is repeated 100 times. The $R_{p}/R_{\ast}$ value is 0.1235$\pm$0.0004 which is 0.32$\%$, 1.0$\sigma$ away from input value (as shown in Fig. \ref{shiLD}). The simulation result is consistent with the real data fitting.

Our simulations and real data modeling with the non-linear law confirm the claims in the literature that a limb darkening law with higher order terms used non-iteratively performs better in deriving transit parameters than laws with lower order terms \citep{Espinoza2015, claret2000, Magic2015, Espinoza2016, Morello2017, Maxted2018, Neilson2018}. The bias propagated to transit parameters is reduced but still exists when the LDC priors are not precisely the same as the real LDCs. 

We also simulate light curves using non-linear law inputs and fitted the light curves with the quadratic law (with and without the non-iterative method). The results are consistent with the results of the quadratic law simulation when the non-linear inputs are set exactly as the non-linear fitting result to real data. However, the simulated light curves are very sensitive to the non-linear law inputs. Any small changes to the LDCs would severely affect the light curves and cause biased fittings with both the non-linear and quadratic laws. Also, the fittings are not biased when we replace the non-linear law inputs with the multimodality values. The multimodality together with the sensitivity (the width of each multi-modality range is narrow) of the input stellar parameters make it very hard and beyond the scope of this work to discuss the reasonable non-linear LDC ranges.

\subsection{Overfitting check}

%The iteration is not convergent to certain values when applying quadratic law to noisier light curves or applying non-linear law LD to WASP-121b TESS level light curves. The other transit parameters derived from the loops after 5 are consistent with input value, as shown in Sec. 3.1, 3.2. The std of the fitting residual after the first a few loops in the iteration is the same as the input noise of the light curves. The simulation indicates that the LD posterior in loops are equivalently accepted by the transit light curves. The LD posterior during the iteration process is not the global optimization but local ones, thus Multimodality. The multimodality appears in LD but not other parameters, e.g. $R_{p}/R_{\ast}$, inclination, semi-major axis.  

%This multimodality does not reveal in single MCMC fitting during the simulation. The multimodality of LD should cause a broader distribution of the posterior in MCMC fitting.The broadening of LD posterior is not happening during simulation due to MCMC difficulty in jumping among the multimodalities as described in Sec. 2.3. The optimization is beyond the purpose of this work. 

%(described in Sec.2.3, 3.1, 3.2)
The multimodality of the LDCs triggers a question: which model does the relation between the transit parameters and light curves with random noise follow, many-to-many or many-to-one? "many-to-many" indicates a robust MCMC process in which the light curves are dominated by transit parameters rather than random noise. "many-to-one" indicates an overfitted model where the fitting is too close to the exact data values regardless of noise \citep{overfitting1995,overfitting2018}. 

We perform a simulation to check if MCMC fitting with the non-linear law is overfitted. We first generate a light curve as described previously. The random noise is with $\sigma$ of 1763ppm. MCMC fitting is performed to the simulated light curve. The parameter values and std of residual are recorded. Next we generate another light curve with the same input as the first but with different random noise (keeping $\sigma$ of random noise the same). We apply the recorded transit parameters to fit the light curve and calculate the std of the residual. The standard deviations of both residuals are 1763ppm. The consistency of the standard deviations indicates that the fitting is not an overfitting.

We perform similar overfitting checks with the
quadratic law with SNR $\sim$5 (LDC multimodality), and $\sim$10 (LDC convergence), and find no evidence of overfitting. We thus conclude that both the quadratic and non-linear laws are not being overfitted.

\section{Summary and Discussion}

We have presented a novel, iterative method using an empirical Bayesian approach to handle limb darkening in modeling the light curves of exoplanet transits. The method is especially effective for the quadratic LD model and works for the non-linear law as well. Prior centers from the synthetic stellar atmosphere modeling are not needed. The method converges to obtain unbiased LDCs. Our method is particularly effective for the case when the LD obtained from the stellar atmosphere model is significantly different from the result obtained from transit fitting. The difference could be due to inconsistency among the stellar atmosphere models (with different geometry or input physics), or from the intrinsic difference among the fitting processes of the stellar atmosphere and transit models \citep{Howarth2011}. \citet{Howarth2011} shows that the difference is nearly a constant offset for the same planet in different wavelengths. Also, the LDC offsets are almost constant or weakly dependent on the stellar effective temperature. This motivates the idea that even if the central values of LDCs are different between the stellar model and transit fitting predictions, the widths of the distributions inferred from the model grids would be unaffected. Pre-knowledge of, e.g., the stellar model prediction, the fitting experience, the observation data quality is inheritable when setting the LDC prior for transit fitting.

We have applied the iterative method to the WASP-121b TESS light curve and compared the parameter estimates with the classic MCMC fitting for both the quadratic law and the non-linear law. The result implies that the iterative method obtains a different, more accurate $R_{p}/R_{\ast}$ at 0.32$\%$, a significance level of 1$\sigma$ compared to the non-iterative method. This improvement is important when taken in the context of exoplanet atmosphere analysis.

Monte-Carlo simulations indicate the iterative method obtains unbiased transit fitting parameters. In WASP-121b simulations, the LDC bias from applying the iterative method is within 0.4$\sigma$. The difference in $R_{p}/R_{\ast}$, inclination and semi-major axis between the iteratively obtained values and the input values is 0.25$\sigma$, 0.35$\sigma$ and 0.66$\sigma$, respectively. By comparison, the non-iterative modeling will cause a bias in $R_{p}/R_{\ast}$ of up to 0.82$\%$ (2.5$\sigma$) when applying shifted LDCs as prior. The simulations also imply that if the real LDCs are as the iterative method obtained, then the non-iterative method obtains biased transit parameters when applying the same prior in the real data fitting. The biased estimate of the transit parameters is the same as the estimate from the non-iterative method fitting to the real data. This hints at the transit parameters of WASP-121b (real data) obtained with the non-iterative method being biased.

We have also presented simulations indicating that the iterative method does not converge for the non-linear law, as well as for the quadratic law with poor signal to noise data, due to the multimodality in fitting the LD law. From the simulation results, we recommend taking the result after the first a few loops in the iteration for the application of non-linear law modeling. The simulations show that applying the non-linear law is better at reducing the bias in parameter fitting compared to the quadratic law when using the non-iterative method.

Our work is a new approach to limb darkening, requiring no theoretical LDCs value from the stellar atmosphere models. We demonstrate the effectiveness of the method with the TESS WASP-121b light curve and simulations. Our proof of concept shows that the iterative method helps solve some of the difficulties in LDCs in transit fitting. Our unbiased fitting results, especially for $R_{p}/R_{\ast}$, are particularly important for the analysis of the exoplanet atmosphere. 
%The exoplanet atmosphere can cause differences in $R_{p}/R_{\ast}$ also at the level of 0.5$\%$ among different observational wavelengths.

%Furthermore, we apply our method to other exoplanets and find that the iteration is not convergent when the host star is in a binary system, e.g. KELT-19Ab. The limb darkening coefficient is not efficient to describe the brightness profile of the host star due to the influence of the companion. The simulation applying parameters of KELT-19Ab but without binary companion presents a convergent iteration procedure. 

\section{acknowledgments}

This work made use of Astroquery\footnote{https://astroquery.readthedocs.io/en/latest/} and the NASA Exoplanet Archive. We would like to thank 
Ranga-Ram Chary for many fruitful discussions, and Juan Carlos Segovia for very useful contributions. We also thank You-Jun Lu for feedback on our work. Fan Yang, Ji-Feng Liu, and Su-Su Shan acknowledge funding from the National Science Fund for Distinguished Young Scholars (No.11425313), National Key Research and Development Program of China (No.2016YFA0400800) and National Natural Science Foundation of China (NSFC.11988101).

%\clearpage
%\newpage
\section{Data availability}
The data underlying this article are available in the article and in its online supplementary material.
\bibliographystyle{mnras}
\bibliography{ref}
\end{document}